# Surface reconstruction and charge modulation in BaFe$_2$As$_2$ superconducting film

S. Kim[1,*,‡], S. Yi[1,*], M. Oh[1,*], B. G. Jang[4], W. Nam[2], Y. -C. Yoo[1], M. Lee[1], H. Jeon[1], I. Zoh[1], H. Lee[1], C. Zhang[1], K. H. Kim[2], J. Seo[3], J. H. Shim[4], J. S. Chae[5,6], Y. Kuk[1,+]

*[1]Department of Physics and Astronomy, Seoul National University, Gwanaku, Seoul 08826, Korea;*

*[2]CeNSCMR & IAP, Department of Physics and Astronomy, Gwanaku, Seoul 08826, Korea;*

*[3]Department of Emerging Materials Science, DGIST, Daegu 42988, Korea;*

*[4]Department of Chemistry, Pohang University of Science and Technology, Gyeongsangbuk-do 37673; Korea*

*[5]Center for Quantum Nanoscience, Institute for Basic Science, Seoul 03760, Korea*

*[6]Department of Physics, Ewha Womans University, Seoul 03760, Korea*

Whether or not epitaxially grown superconducting films have the same bulk-like superconducting properties is an important concern. We report the structure and the electronic properties of epitaxially grown Ba(Fe$_{1-x}$Co$_x$)$_2$As$_2$ films using scanning tunneling microscopy and scanning tunneling spectroscopy (STS). This film showed a different surface structure, $(2\sqrt{2} \times 2\sqrt{2})$R45º reconstruction, from those of as-cleaved surfaces from bulk crystals. The electronic structure of the grown film is different from that in bulk, and it is notable that the film exhibits the same superconducting transport properties. We found that the superconducting gap at the surface is screened at the Ba layer surface in STS measurements, and the charge density wave was observed at the surface in sample in the superconducting state.

*: Equal contribution

+: Corresponding author: ykuk@phya.snu.ac.kr

‡: Present address: Center for Nano Science and Technology, NIST, Gaithersburg MD 20899, USA

PACS numbers: 74.20.Rp, 74.78.-w, 74.70.-b, 07.79.Cz

# I. INTRODUCTION

Many high transition-temperature ($T_C$) superconductors have been studied using bulk-grown samples because it is difficult to grow materials with multiple components. Recently, high-quality ternary or quaternary superconductor films have been successfully grown by multi-component growth methods.[1] Moreover, the introduction of pulsed laser deposition (PLD) to grow superconducting films has enabled layer-by-layer control on various high-$T_C$ superconducting thin films.[2-4] These grown films are known to have the same crystal structures and similar transport properties as bulk-grown superconductors. However, the surface structures in the films may be different from those in as-cleaved bulk samples. In addition, if the film is very thin, the quantum bound state effect is manifested,[2] or superconductor-to-insulator phase transition occurs if there are impurities or a magnetic field is applied.[6–8] In this paper, we report systematic studies to determine whether the surface structures and transport properties of these thin films are identical or different to the counterpart of bulk-grown specimens.

Following the discovery of iron pnictide superconductors with relatively high $T_C$,[9] much research has been conducted on their crystal structure, transport properties,[2,10] electronic structures and corresponding pairing states, and the role of antiferromagnetism on superconductivity.[11–13] Recently, resonant pairing scattering between multi-bands on iron pnictide superconductors has been investigated by angle resolved photoemission spectroscopy (ARPES),[14,15] scanning tunneling microscopy (STM) and scanning tunneling spectroscopy (STS) experiments.[16–19] Those studies were performed on surfaces of cleaved bulk crystals,[20–26] and doped $Æ Fe_2As_2$ (one alkali earth, two irons, two arsenics) superconductors were the ones most commonly studied. It is relatively straightforward to synthesize $Æ Fe_2As_2$ superconductors from a bulk single crystal and to control the doping level, and it is easy to cleave the surfaces. Most measurements using ARPES, STM and STS have been performed on the cleaved surfaces, where the alkali earth planes were exposed. It has been assumed that the electronic structures measured on the cleaved specimen are equivalent to those of the bulk material.

Quaternary superconductors have been grown using the multi-component molecular-beam-epitaxy method but only a few research groups have succeeded to grow high-quality films.[1] The alternative is PLD growth from a premade polycrystalline target filled with a buffer or a reactive gas. For PLD-grown specimens, atomic structures have been studied by transmission electron microscopy (TEM) and the superconducting transition by transport measurements.[27–30] However, the electronic structure on the surfaces and the influence of temperature on the superconducting order parameters have not yet been studied extensively. It is also noteworthy that the surface structures of an $Æ Fe_2As_2$ cleaved at room temperature and cryogenic temperature are predicted to be different, because the surface atoms can find different equilibrium positions at room temperature. For example, for a $Ba(Fe_{1-x}Co_x)_2As_2$ sample, the surfaces exhibit a $(2 \times 1)$ or a $(\sqrt{2} \times \sqrt{2})R45º$ structure when cleaved at low temperature; however, only a $(\sqrt{2} \times \sqrt{2})R45º$ structure is observed on surfaces cleaved at an elevated temperature.[25,26,31,32]

In this study, we investigated the differences in the surface structures and the physical properties of a quaternary material, $Ba(Fe_{1-x}Co_x)_2As_2$ (x=0.08) (BFCA), grown using PLD under ultrahigh vacuum (UHV) environment, compared with a surface cleaved from bulk-grown specimen, by using STM/STS measurements. We observed an entirely different structure, $(2\sqrt{2} \times 2\sqrt{2})R45º$, on the surface and superconducting gap ($\Delta$) variation at various planes and at various temperatures. In addition, we observed charge modulation on the Ba-terminated surface.

## II.    EXPERIMENTAL PROCEDURE

Electron-doped Ba(Fe$_{1-x}$Co$_x$)$_2$As$_2$ (x = 0.08) pellets with $Tc$ = 26 K were used as the targets for PLD growth. The targets were fabricated using the conventional solid-state reaction method. A 0.05-wt% Nb-doped SrTiO$_3$(100) (STO) wafer was used as the substrate.[33,34] Before growing the superconducting film, the STO wafer was etched with buffered hydrofluoric acid (HF) and annealed at 1100 °C to yield a clean TiO$_2$-terminated surface.[35] Figure 1a shows the low energy electron diffraction (LEED) pattern of the $(1 \times 1)$ peaks in the STO substrate. PLD growth was conducted using a KrF excimer laser (wavelength 248 nm, pulse duration 20 ns) without buffer gas under UHV condition (base pressure $< 2 \times 10^{-10}$ Torr). During the growth, the temperature of the sample was maintained at 700 ± 25 °C, *i.e.*, within a narrow temperature window. The film growth was monitored using reflection high energy electron diffraction (RHEED).[27,36,37] The film thickness was varied from 1 to 100 monolayers (MLs). Although the difference in the lattice constant between STO and BFCA is small (3.90 and 3.96 Å, respectively), the initial three MLs did not exhibit good layer-by-layer growth, as judged by the RHEED pattern (not shown here). We could observe neither superconducting gaps in the STS measurement nor superconducting transition in four-probe transport measurement in the films with ≤ 3 MLs.

When the growth was completed, the film crystallinity was confirmed through a LEED pattern and the sample was transferred to the STM, which is connected via UHV extension to the growth chamber. STM measurement was performed at ~ 7 K under UHV. A polycrystalline iridium tip was used after chemical etching followed by field ion beam (FIB) sharpening. A UHV suitcase at the established pressure (*i.e.*, below $2 \times 10^{-10}$ Torr) was used to transfer the sample from the growth chamber to another STM operated at 4.3 K.

### III.    SURFACE STRUCTURE

Figure 1b shows the LEED pattern of a 10-ML-thick BFCA sample, in which a $(2\sqrt{2} \times 2\sqrt{2})$R45° structure appeared unexpectedly. Figure 1c is an STM image of the 10-ML-thick BFCA surface on STO. The first three layers show island growth (not shown here), and following 4–10 MLs show step-flow growth, as shown in this 135 nm × 135 nm image. Note that screw or edge dislocations were often observed in the STM images. In the initial island growth mode, the nucleation centers are scattered irregularly with an average separation of 30–70 nm. Domain boundaries appear between neighboring grains as indicated by green arrows in Fig. 1c. Layers of four different heights are visible on the 10-ML surface. In the magnified STM image as shown in Fig. 1d, atomic resolution can be achieved on the BFCA surfaces. The protruding blobs, which appear as single atoms, are spaced 1.12 nm apart and are regularly arranged. These blobs are not single Ba atoms but, rather, clusters of four atoms, as will be discussed at the end of this section.

For a cleaved BFCA sample, it is known that a surface half covered by Ba atoms is energetically more stable than a surface fully covered with 1 ML of Ba atoms. Moreover, the half-covered surface can satisfy the charge neutrality.[38] There are two structures that satisfy this criterion. One is a structure in which every alternate Ba row at 3.96 Å separation is missing, yielding a $(2 \times 1)$ structure with 7.92-Å separation between neighboring rows, as shown in Fig. 2a. It has been reported that the surface cleaved at the cryogenic temperature shows this structure. The second case is where every alternate diagonal row is missing, yielding a $(\sqrt{2} \times \sqrt{2})$R45º structure, as shown in Fig. 2b. The spacing in this case is 5.6 Å. This structure was observed at the room-temperature cleaved surface. Figure 2e shows the $24.5 \times 24.5$ nm topography of a 10-ML PLD-grown BFCA surface. The spacing between neighboring blobs is 11.2 Å, suggesting a $(2\sqrt{2} \times 2\sqrt{2})$ R45º structure (as shown in Fig. 2c), which differs significantly from the structure of the cleaved surfaces. Many missing rows are visible and the spacings between neighboring rows are either 16.8 or 22.4 Å, as shown schematically in Fig. 2d. Further, the upper terrace in Fig. 2e is quite disordered, which occurs when the local Ba content is less than 0.5 ML, yielding missing rows or disorder. If the original Ba content on the surface in the as-grown sample was 0.5 ML, the local Ba deficiency would be matched by greater number of Ba atoms in some other areas. The increased Ba content can be seen in the excess Ba clusters at the island of Fig. 2f. Figure 2g shows a higher-resolution image of missing rows and of fragments of Ba atoms near these missing rows. As indicated by the arrows, smaller blobs are apparent near missing rows and defects. The regularly appearing blobs in Figs. 2e and 2f may be composed of multiple Ba atoms. This superstructure can be attributed to the fact that the surface cleaved at low temperature has a surface energy different from that produced at the growth temperature during the epitaxial growth process. In addition, the locally lower or higher Ba densities can be explained by considering the diffusion-limited kinetics in the PLD growth process at the growth temperature.

Figure 3 illustrates the mechanism through which these reconstructed surfaces are formed. Figures 3a and 3b show STM topography results obtained at the sample bias (*V*) of - 0.5 V and +0.1 V, respectively. The reconstructed lattice generally appears in a blob shape, as shown in Fig. 3a, but exhibits a ring shape at a specific bias, as shown in Fig. 3b. Careful observation of the ring shape in Fig. 3b indicates that each ring appears to be a cluster of four Ba atoms. There could be two possible scenarios. First, it can be assumed that four Ba atoms are shifted to the center on a uniformly distributed $(\sqrt{2} \times \sqrt{2})$ surface to create each bright site. Second, one can conclude that half the Ba atoms from the fully occupied Ba $(1 \times 1)$ surface remain, as shown in Fig. 3d, creating a $(2\sqrt{2} \times 2\sqrt{2})R45^{\circ}$ reconstructed surface. Using density functional theory (DFT), we calculated the energy stability of these two cases. As a reference structure, we consider a slab structure with fully occupied Ba and As surfaces in each side of the slab. In the first case (Fig. 3c), the energy difference per unit cell is -0.254 eV, which indicates higher stability than the second case (Fig. 3d), for which a value of -0.171 eV is obtained. Through STM image analysis and DFT calculations, we can conclude that the surface of epitaxially grown BFA film has a $(2\sqrt{2} \times 2\sqrt{2})R45^{\circ}$ superstructure with the unit of a cluster of four Ba atoms centered to each other.

## IV.  TRANSPORT PROPERTIES AND ELECTRONIC STRUCTURE MEASUREMENT

For doped superconductors such as BFCA, effective annealing yields increased $T_C$ and critical current density ($Jc$).[39,40] As a result, it is understood that inhomogeneity of the Co doping level occurs at nanometer scale. Figure 4 shows the change in resistance with temperature. Figures 4a and 4b show the *ex-situ* and *in-situ* measurement results obtained with four-point probes substituted in place of an STM probe, respectively. The *ex-situ* result shows an onset $Tc$ of 22 K and zero resistivity at 21 K. This $T_C$ is slightly lower than that measured on a PLD target. The *in-situ* result shows an onset $T_C$ of 24 K and zero resistivity at 22 K. For many repeated measurements, variations of these temperatures are observed due to the local variation of doping. In Fig. 4b, the temperature dependence of the resistivity does not follow that of metal immediately above the onset $T_C$, because a very low level of Nb doping in the STO substrate was used in this study.

To date, the reported results of point STS performed on cleaved BFCA surfaces have been dominated by the V-shape of the d-wave type superconducting gap with unclear coherence peaks.[32] A recent ARPES measurement and DFT study have suggested that an exposed Ba layer screens the superconducting property of the As-Fe-As layer.[24] In this study, we also observed the screening effect by the Ba surface, by measuring the layer-dependent tunneling spectrum. Figure 5a is an STM topographic image of a BFCA film surface grown to 10 MLs. In most areas, well-ordered $(2\sqrt{2} \times 2\sqrt{2})$R45º reconstructed structures are observed, but occasionally, areas such as "B" in Fig. 5a are apparent. In those areas, the surface Ba atoms are removed and the As-Fe-As trilayer is exposed. Analysis of the height profile determined from the STM topography indicates that the step height is generally 12.8 Å, which is the unit cell height, or ~ 6.4 Å, which is half of the unit cell height. However, the depression height of this area "B" is less than 6.4 Å. As shown in Fig. 5b, in the outer region covered by Ba atoms, the tunneling spectrum exhibits a V-shaped gap feature, which is similar to previous results for the cleaved surface.[32,39] On the other hand, the tunneling spectrum in region "B" shows a more pronounced superconducting gap feature with $\Delta \approx 6$ meV and clearer coherence peaks than those in the Ba-covered region. This result clearly implies that the electronic state of the surface Ba atoms screens the superconducting property of the As-Fe-As trilayer to the STM tunnel junction. We confirmed that the As-Fe-As layer is superconducting and that it is not easy to directly measure superconductivity on the surfaces of $\mathcal{Æ}Fe_2As_2$ iron pnictide superconductors because of the surface $\mathcal{Æ}$ atoms. Figure 5c shows the evolution of $\Delta$ with temperature on the surface with exposed As-Fe-As trilayers.

The formation of the superconducting state can be clearly determined by measuring point tunneling spectra; however, the quasiparticle dynamics related to the superconducting pairing mechanism cannot be inferred. Instead, measurement of quasiparticle interference (QPI) patterns provides informed clues for the pairing symmetry in cuprate superconductors and the electronic nematicity.[17,41–43] In this technique, STS spectra at different positions are recorded and the differential conductance map (local density of states map) at a given energy can be obtained by collecting STS signal at all positions. Fourier transformation of the position-dependent local density of states yields the quasi-particle interference (QPI) pattern in $\mathbf{k}$-space. This method is very useful if the As layer is only exposed. However, the QPI patterns obtained on the Ba-terminated surface seem to reveal different QPI behavior. Figure 6 shows QPI patterns obtained at $V = 7$ and 15 meV. Unlike the QPI results obtained on the As-terminated cleaved surface, quasiparticle interference between hole and electron pockets is not observed. Instead, quite unusual higher order spots associated with charge modulation on the Ba-terminated surface are observed.

By analyzing the electronic structure with different energies, it was observed that the topography images (and conductance maps) exhibited different patterns as $V$ approached the vicinity of the superconducting gap. Figure 7 shows the changes in the $V$-dependent topography (upper) and fast Fourier transformation (FFT) images (lower). As apparent from the changes in the topography images, the surface exhibits a regular checkerboard pattern at high $V$; however, the four-fold symmetry seems to be broken in the vicinity of the superconducting gap, and the blob shape changes to ring shape, as shown in Fig. 7e (or Fig. 3b). This tendency is also reflected in the FFT results, with some peaks other than the $(2\sqrt{2} \times 2\sqrt{2})$ R45° reconstruction peaks appearing in accordance with the changes in topography. In fact, this charge modulation is similar to the findings of a previous study on a $(\sqrt{2} \times \sqrt{2})$R45° surface.[26] This electronic structure change on the surface may resemble the charge density wave reported on the surface of a cleaved cuprate high-$T_C$ superconductor.[44]

Figure 8 shows an analysis of the energy-dependent changes in several peaks shown in Fig. 7. Figure 8a shows the positions of the peaks showing significant changes in Fig. 7e. The (1, 0) and (0, 1) peaks appear due to $(2\sqrt{2} \times 2\sqrt{2})$R45° reconstruction, and the (1, 1) peaks appear strong in the 50–200 mV region, where each Ba atom site is resolved as the lattice site becomes ring-shaped. The topographic image appears to have two-fold symmetry rather than a complete four-fold symmetry that can be interpreted as tip asymmetry. It is apparent that the peaks labeled A $\left(\frac{1}{6}, \frac{1}{3}\right)$ and B $\left(-\frac{1}{2.5}, \frac{1}{3}\right)$ appear bright at a specific bias, and these peaks represent charge modulation over a longer range than the reconstructed lattice. Figure 8b is a plot of the intensity changes of these peaks with respect to $V$. To facilitate comparison, the value divided by the intensity of the $(0, 1)$ peak was taken and the change in the $(1, 0)$ peak was shown as a reference. As shown in Fig. 8b, these peaks appear strongly at 0–100 mV and decrease in an outward direction. This change in the energy-dependent charge modulation is similar to that of a charge density wave; therefore, this result may indicate the possibility of a charge density wave transition that has not been previously reported for an iron-based superconductor.

## V. CONCLUSION

In this study, we performed STM/STS measurements on a $Ba(Fe_{1-x}Co_x)_2As_2$ (x=0.08) surface epitaxially grown via PLD. We observed a $(2\sqrt{2} \times 2\sqrt{2})R45^o$ surface reconstruction on the epitaxially grown BFCA film, which was not measured on the cleaved surfaces. We confirmed the electronic screening effect of the surface atoms through layer-dependent tunneling spectrum analysis. In addition, we analyzed the changes in the electronic structure to observe the charge modulation on the surface; the results indicate that this behavior may be similar to that of a charge density wave in a cuprate superconductor. This work will serve as the foundation for future STM and STS studies on epitaxially grown 122 compound films, and will provide clues towards a more profound understanding of the superconducting mechanism in iron-based superconductors.

Acknowledgements

This work is supported in part by the National Research Foundation of Korea (NRF) Grant (NRF-2006-0093847, NRF-2010-00349). The work at CeNSCMR was supported by the National Creative Research Initiative (2010-0018300) program by NRF of South Korea. JSC was supported by the IBS research program IBS-R027-D1.

**FIGURE CAPTIONS**

Fig. 1: (a) LEED pattern of clean STO substrate at 70 eV. (b) LEED pattern of 10-ML Ba(Fe$_{1-x}$Co$_x$)$_2$As$_2$ (BFCA) samples at 70 eV. (c) STM topography of 10-ML BFCA surface obtained on 135 nm × 135 nm area at $V_{tip}$ = 2 V, $I_{tun}$ = 30 pA. (d) STM topography of 10-ML BFCA surface obtained on the 25.5 nm × 25.5 nm area at $V_{tip}$ = − 2 V, $I_{tun}$ = 20 pA. Green arrows in (c) and (d) point the nucleation sites

Fig. 2: In (a)–(d), the small gray dots and green filled circles indicate the Ba atom positions in the bulk lattice and Ba atoms on surface, respectively. (a) $(2 \times 1)$, (b) $(\sqrt{2} \times \sqrt{2})$R45º, and (c) $(2\sqrt{2} \times 2\sqrt{2})$R45º Ba-terminated surface of BFCA, respectively. (d) Missing row structure in $(2\sqrt{2} \times 2\sqrt{2})$R45º structure. (e) 24.5 nm × 24.5 nm topography of 10-ML BFCA samples. $V_{tip}$ = − 2 V, $I_{tun}$ = 30 pA (f) 56 × 56 nm topography of 10-ML BFA samples. $V_{tip}$ = − 2 V, $I_{tun}$ = 30 pA (g) 29 × 29 nm topography of 10-ML BFA samples. $V_{tip}$ = − 2 V, $I_{tun}$ = 50 pA. Green arrows indicate smaller blobs apparent near missing rows and defects.

Fig. 3: (a)–(b) Topographic images of (a) 10 nm × 10 nm area obtained at $V_{tip}$ = + 0.5 V, and (b) 3.3 nm × 3.3 nm area obtained at $V_{tip}$ = + 0.1 V. (c)–(d) Atomic configurations of Ba-covered surface. (c) In $(\sqrt{2} \times \sqrt{2})$R45º structure, four Ba atoms shift close to the center, forming a $(2\sqrt{2} \times 2\sqrt{2})$R45º structure. (d) A second model to explain the $(\sqrt{2} \times \sqrt{2})$R45º structure; half of the Ba atoms are selected and occupied from the $(1 \times 1)$ surface.

Fig. 4: Temperature dependence of resistance of PLD-grown BFCA samples, measured (a) *ex-situ* and (b) *in-situ*, respectively.

Fig. 5: (a) Topographic image of PLD-grown BFCA surface at $V_{tip}$ = − 2 V, $I_{tun}$ = 30 pA. The exposed FeAs area is indicated by "B". (b) Tunneling spectra of marked positions in (a). (measured at 4.5 K) (c) Gap evolution with temperature. Black, blue and red lines were measured at 9.2 K, 12 K and 15 K, respectively.

Fig. 6: Quasiparticle interference patterns obtained at $V =$ (a) 7 and (b) 15 meV, respectively.

Fig. 7: (upper) $V$-dependent topographic images (10 nm × 10 nm) and (lower) FFT images obtained at (a) − 500, (b) − 50, (c) − 5, (d) + 5, (e) + 50, (f) + 500 mV, respectively.

Fig. 8: (a) Magnified view of FFT image of Fig. 7e. (b) Intensity plot of varying spots with respect to $V$. The marked points are indicated in (a). Offset is applied for convenience.

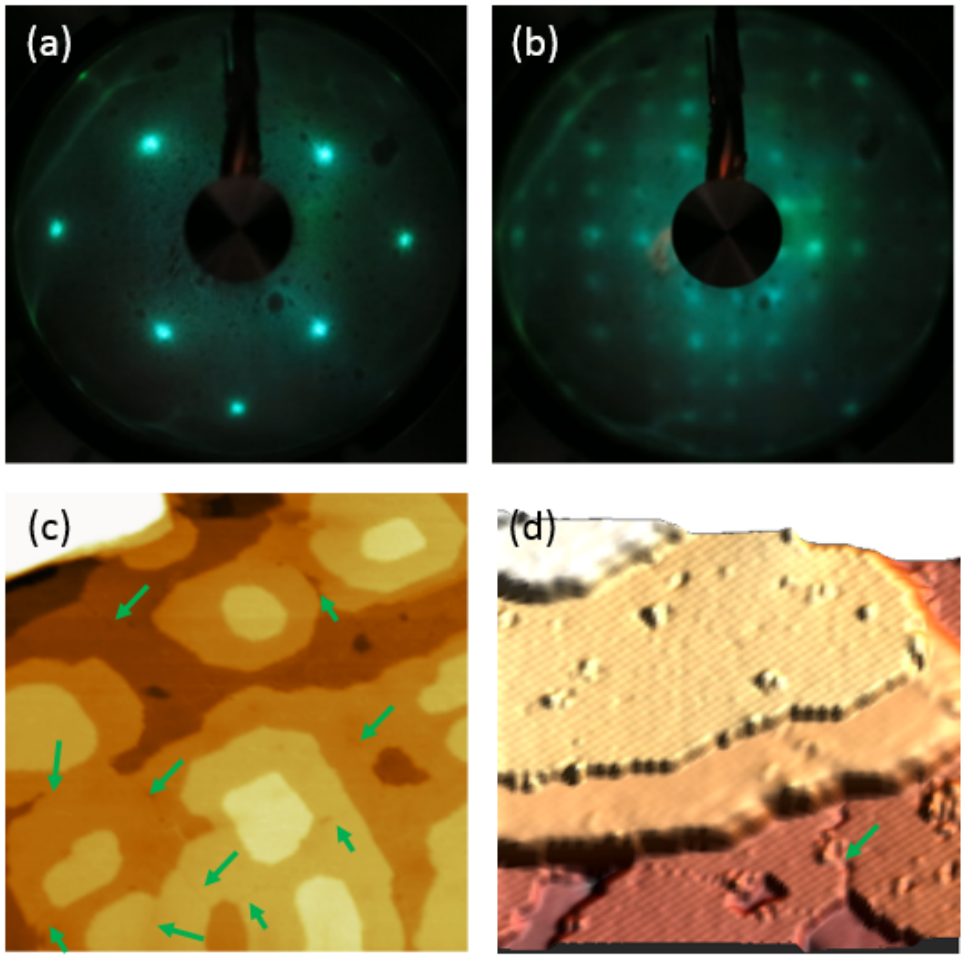

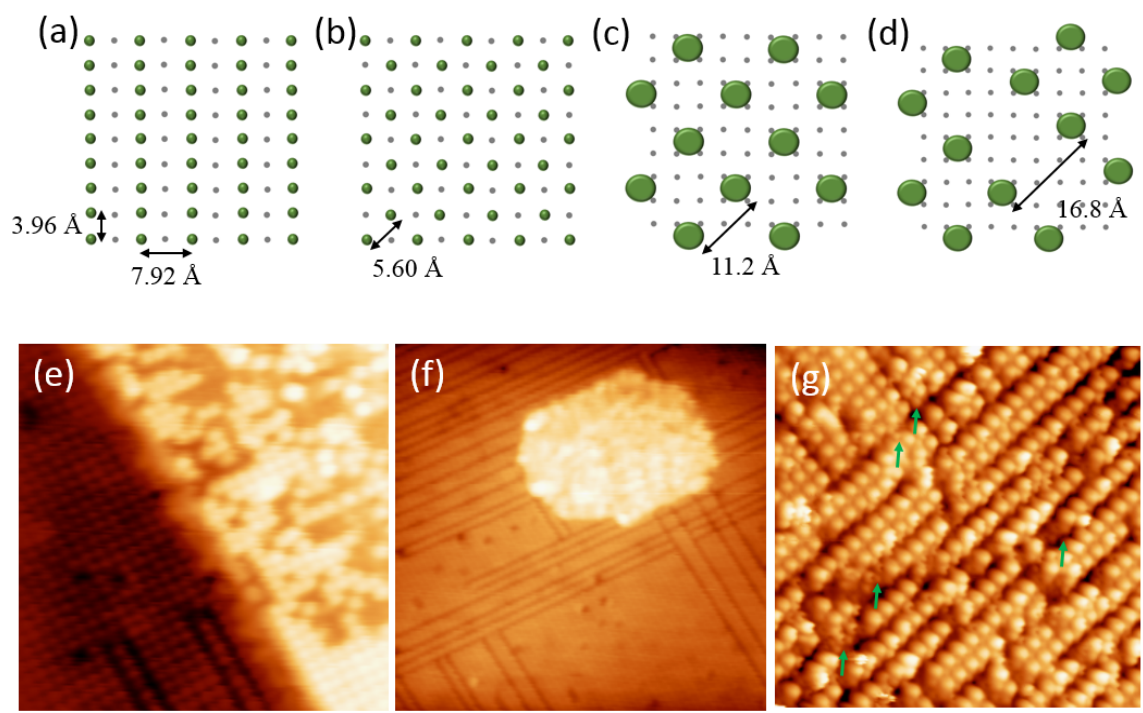

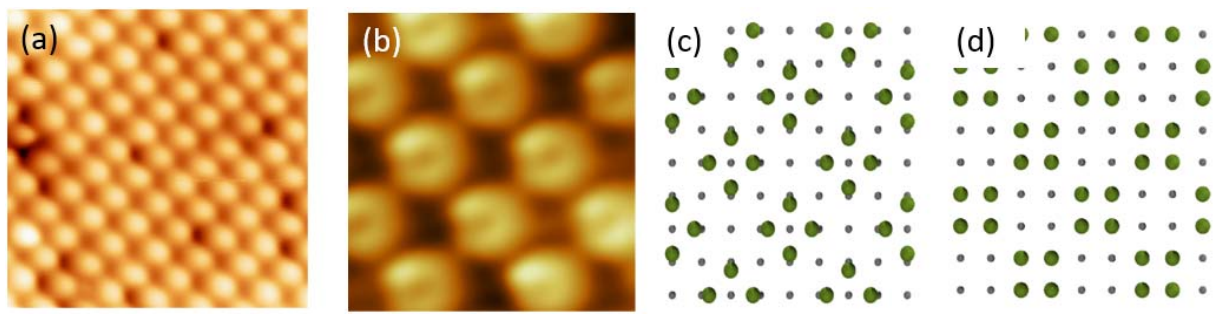

Kim Fig. 3

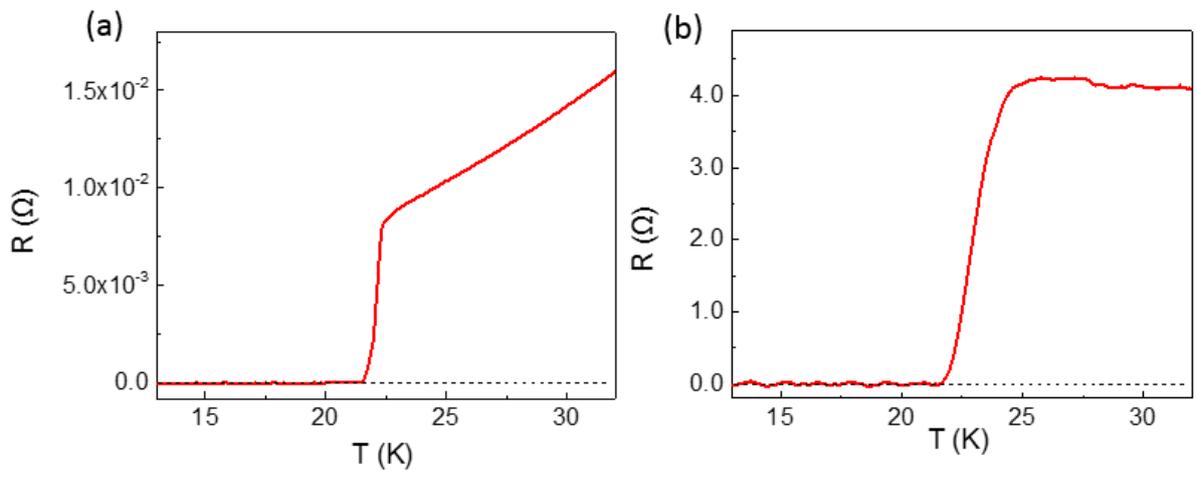

Kim Fig. 4

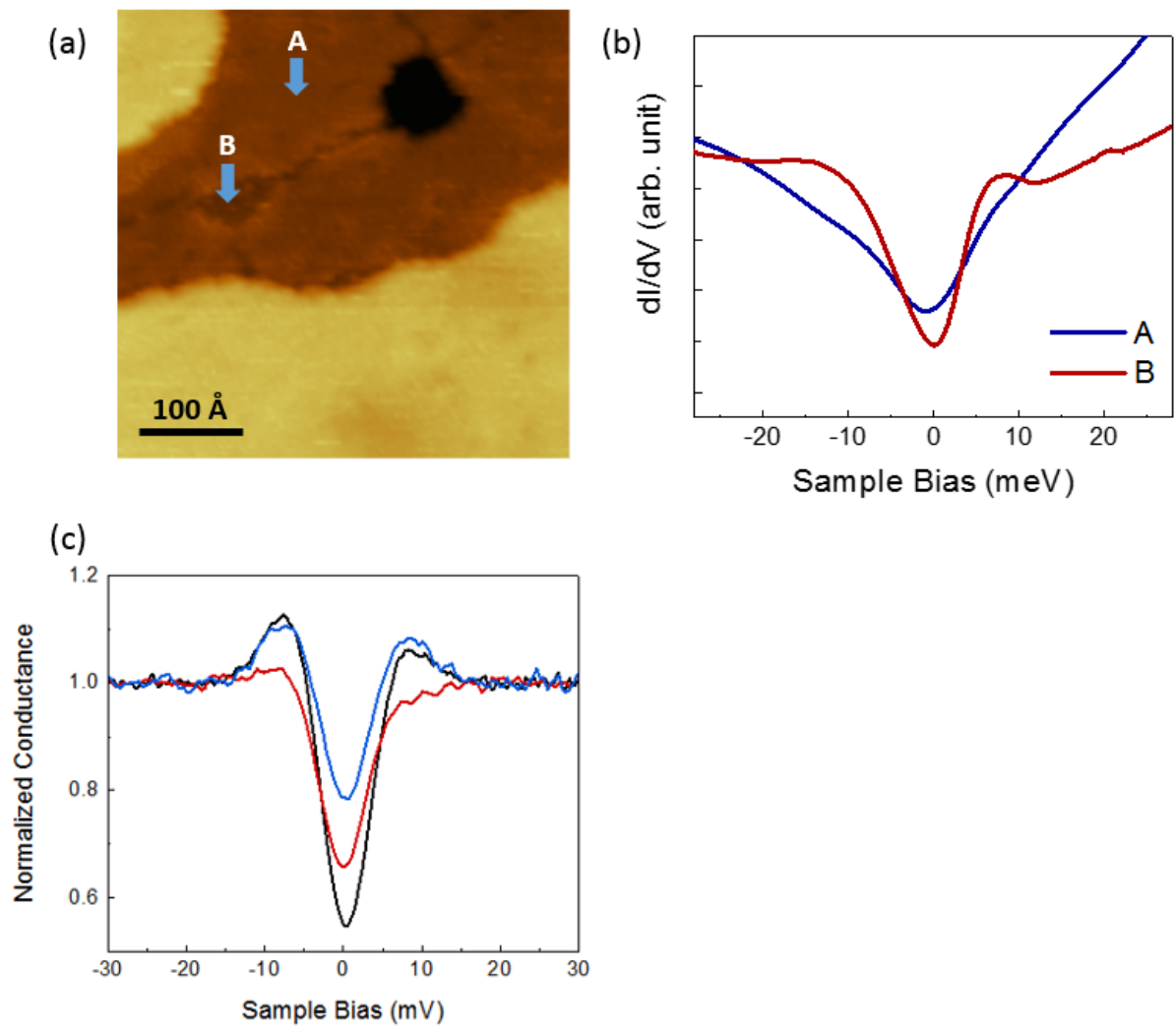

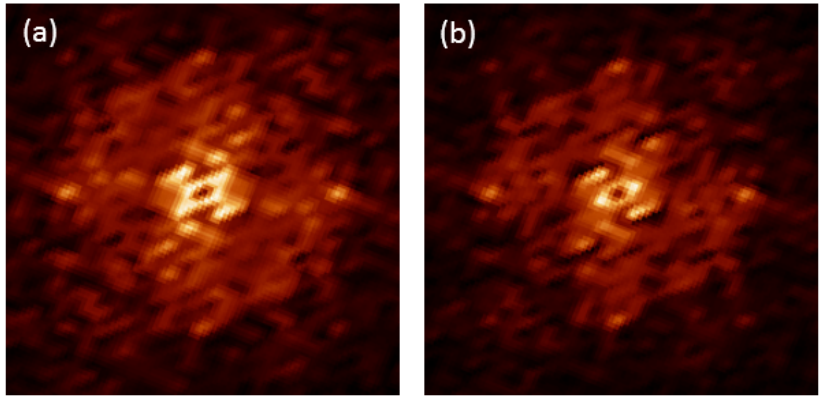

Kim Fig. 6

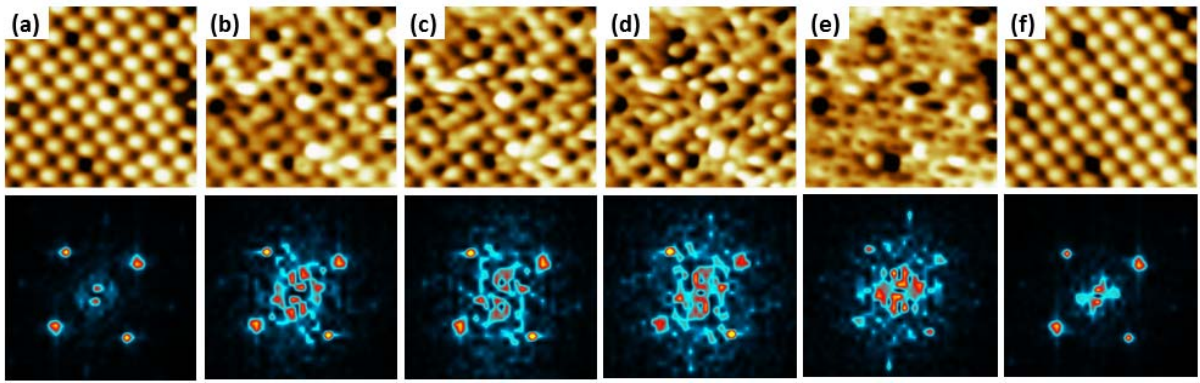

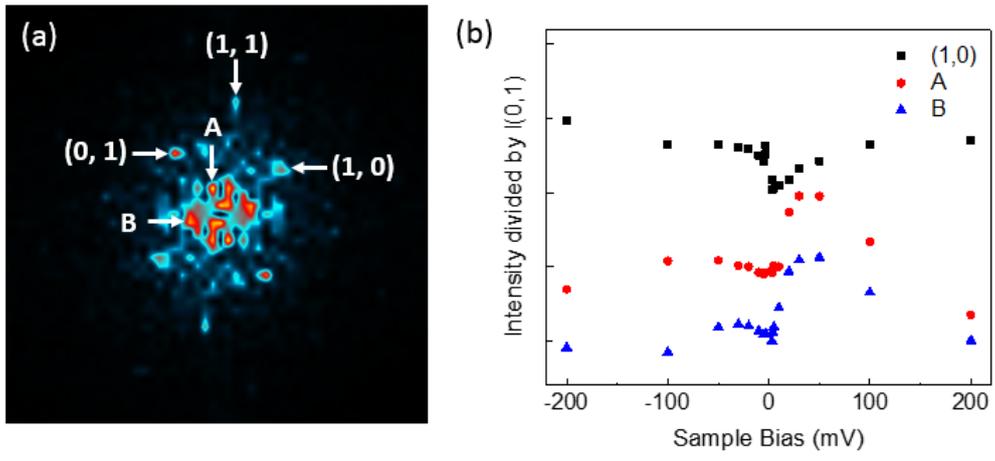